\documentclass{pasa}

\jid{PASA}
\doi{10.1017/pas.2016.xxx}
\jyear{2016}

\title[The MORX Catalogue]{The Million Optical - Radio/X-ray Associations (MORX) Catalogue}
\author[E. Flesch]{\textbf{Eric W. Flesch }$^{A,B}$\\
\\
\affil{$^A$PO Box 5, Whakatane, New Zealand}
\affil{$^B$Email: eric@flesch.org}}

\begin{document}

\begin{abstract}
This automated catalogue combines all the largest published optical, radio and X-ray sky catalogues to find probable radio/X-ray associations to optical objects, plus double radio lobes, using uniform processing against all input data.  The total count is 1\,002\,855 optical objects so presented.  Each object is displayed with J2000 astrometry, optical and radio/X-ray identifiers, red and blue photometry, and calculated probabilities and optical field solutions of the associations.  This is the third and final edition of this method.   
\end{abstract}


\maketitle

\section{Introduction}

Astronomy is a large science with many areas of specialty, so few researchers publish across many of the subdisciplines at once.  Perhaps this limitation delayed publication of joined optical, radio and X-ray sky data in the early 2000's when {\it ROSAT} X-ray and NVSS \& FIRST radio source catalogs were already available.  In 2004 Martin Hardcastle and I endeavoured to fill the gap by publishing the Quasars.org catalogue (QORG: Flesch \& Hardcastle \shortcite{QORG}); this used a uniform data-driven algorithm to align and overlay the optical, radio and X-ray input data while still accomodating the peculiarities of each input catalogue.  This was updated in 2010 by the Atlas of Radio/X-ray Associations (ARXA: Flesch \shortcite{ARXA}) with \textit{Chandra} and \textit{XMM-Newton} X-ray data and larger FIRST \cite{FIRST} radio data.   

Today the dynamic databases like NED\footnote{NASA/IPAC Extragalactic Database, http://ned.ipac.caltech.edu} and SIMBAD\footnote{SIMBAD database at CDS, http://simbad.u-strasbg.fr/simbad} collect all those data and present them with their originally published astrometry.  While this resource is essential, it omits the astrometric refinement made possible by optical field solutions (where X-ray fields are aligned to their optical background) and leaves the determination of causal association (i.e., that a particular optical object is the true source of nearby radio/X-ray detections) to the user's devices.  Thus the approach of QORG and ARXA is still needed as an aid to astronomers' quick selection of objects of interest.  

Since ARXA's publication in 2010 there have been continued releases of \textit{XMM-Newton} and \textit{Chandra} catalogued data, and concluding releases from the FIRST project.  Also the first catalogue of \textit{Swift} X-ray sources has been published, and multiple identification releases from the prolific SDSS\footnote{Sloan Digital Sky Survey, http://sdss.org} project, and individual publications.  Equally important for ARXA-style processing is the quality of the optical background data, and since 2010 I have added the USNO-B \cite{MONET} optical data, plus much SDSS optical data, to make an optical background of over a billion objects with very few one-colour objects compared with before.  Combining all these data using the QORG/ARXA processing now results in over a million optical objects with radio/X-ray association of 40\%-100\% likelihood of being true.  Thus, this seems the right time to publish this third edition as the ``Million Optical - Radio/X-ray Associations Catalogue'' (hereafter: MORX), which is the final edition of this method.

This catalogue is available in both flat file and FITS formats\footnote{at http://quasars.org/morx.htm}, with a ReadMe.  Table 1 shows a few sample lines of the flat file with some explanation of the columns, but the ReadMe gives full details of the layout and contents.  Figure 1 shows the sky coverage of MORX.

\begin{figure*} 
\includegraphics[scale=0.5, angle=0]{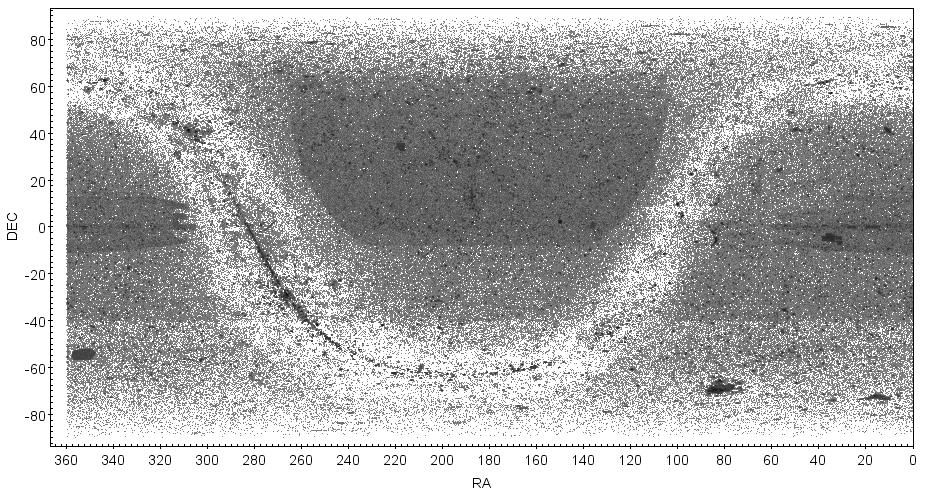} \\
\tiny{Chart produced with TOPCAT \cite{TAYLOR}.}  
\caption{Sky coverage of the MORX catalogue, darker is denser.  Large dense places represent FIRST coverage.  The density boundary at declination -40$^{\circ}$ shows the southern edge of NVSS coverage.} 
\end{figure*}

\begin{table*} 
\caption{Sample lines from the MORX catalogue (left half placed on top of right half)} 
\includegraphics[scale=0.325, angle=0]{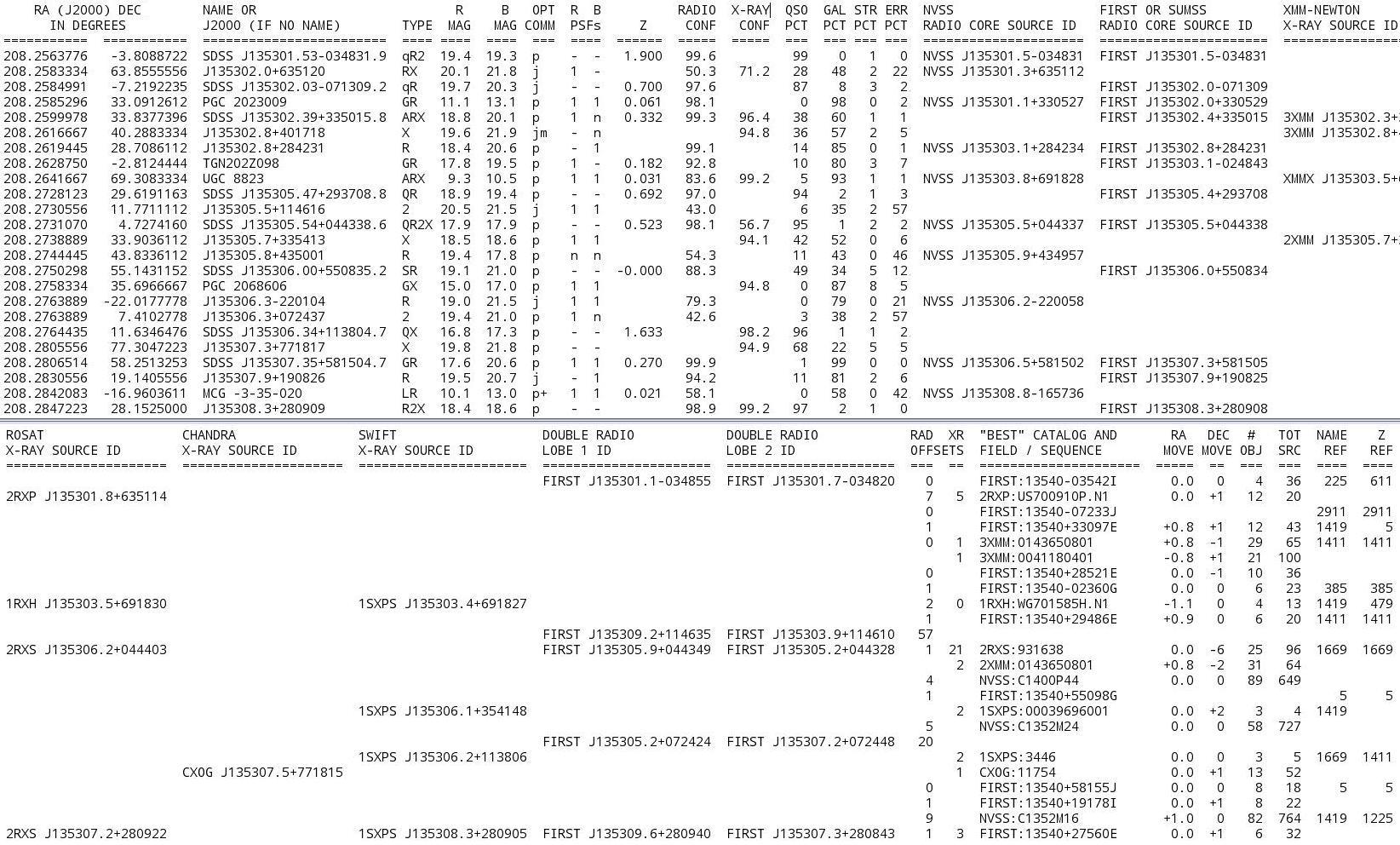} 
\tiny 
Notes on columns (see ReadMe for full descriptions): 
\begin{itemize}
	\item TYPE:  R=core radio, X=X-ray, 2=double radio lobes, Q=QSO, A=AGN, q=photometric quasar, G=galaxy, L=LINER, S=star. 
  \item OPT COMM:  comment on photometry: p=POSS-I magnitudes, so blue is POSS-I O, j=SERC Bj, +=optically variable, m=nominal proper motion.
  \item R/B PSFs:  '-'=stellar, 1=fuzzy, n=no PSF available, x=not seen in this colour.
  \item RADIO/X-RAY CONF:  calculated percentage confidence that this radio/X-ray source is truly associated to this optical object. 
  \item QSO/GAL/STR PCT:  based on its photometry and the association(s), the calculated percentage confidence that this optical object is a QSO/galaxy/star. 
  \item ERR PCT:  calculated percentage chance that this association is wrong, equals 100\% minus the combined radio/X-ray confidence.
  \item RAD/XR OFFSETS:  in arcsec, the astrometric offset from the optical object to the best (i.e., highest confidence associated) radio/X-ray source, after that source has been shifted by its optical field solution.
  \item ``BEST'' CATALOG AND FIELD / SEQUENCE:  identifies the radio/X-ray field used in the optical field solution.
  \item RA MOVE:  in arcsec, the East-West shifts of the optical field solution pertaining to this source, calculated in integer RA arcsecs, i.e., 1/3600$^{th}$ of a RA degree. 
  \item DEC MOVE:  in arcsec, the North-South shifts of the optical field solution pertaining to this source, calculated in integer arcseconds. 
  \item \# OBJ:  the number of sources which were associated to optical objects (with $>$70\% confidence) by this optical field solution.
  \item TOT SRC:  the total number of sources in this radio/X-ray field.
  \item REF \& ZREF: citations for name and redshift; citations are indexed in the file "MORX-references.txt".
\end{itemize} 
The full table can be downloaded from http://quasars.org/morx.htm, also available in FITS.
\end{table*}

The QORG paper presents the full details of the matching techniques in its appendix A, but in the sections below I summarize the method and issues involved in constructing this catalogue, including any updated techniques, and give a brief roundup of all the optical, radio and X-ray source catalogues.  Thus this paper gives a comprehensive overview suitable for most users of the catalogue.

\section{The Optical Background used in MORX}

Assembly of the background optical data pool was documented in QORG \cite{QORG} as consisting of red \& blue magnitudes and PSFs from Cambridge Automatic Plate Measuring (APM) \cite{MI} data and United States Naval Observatory (USNO-A) \cite{MONET98} data.  Those two projects used purpose-built scanners to read glass copies of first-epoch National Geographic-Palomar Observatory Sky Survey (POSS-I) $E$ and $O$ plates, and the UK Schmidt Telescope Sky Survey (UKST) ESO-R and SERC-J plates.  That data amounted to 671M objects and was used in QORG and ARXA \cite{ARXA} which recalibrated the magnitudes as detailed in QORG. 

The present catalogue features a very significant improvement in that the optical USNO-B \cite{MONET} catalogue, which includes data from the second-epoch POSS-II which is deeper than POSS-I, has been added.  This brings the optical background to 933M objects.  Also, importantly, one-colour objects (red-only or blue-only) are much reduced and a class of POSS-I one-colour objects were identified as invalid and 12M deleted.  Figure 2 compares the before-and-after using a 15$^{\circ}$-sq tile of sky, the shadings showing object density.  

The top panel of Figure 2 reveals the rectangular boundaries of the POSS-I plates (cropped and fitted to eachother); the edges become visible because the de-duplication algorithm conserved unique objects from each plate (missed by the other plate) in the overlapping strip, plus duplicates $>$3 arcsec apart survived the de-duplication, so the edges have extra density.  Some plate names are annotated.  The plates are seen to have higher optical density at their centres than toward the edges, but those extra objects in the plate centres are spurious.  The problem was in the striving for completeness: the glass copies of the POSS-I plates used by APM, and the APM deep scanning of those glass copies, both endeavoured to capture the faintest optical imagery but the combined result was the inclusion of exposure artefacts as what could whimsically be called ``carbon stars''.  Those one-colour spurious objects have now been deleted from my optical background on the criterion that they were unmatched to the deeper POSS-II data from USNO-B.

The bottom panel of Figure 2 shows the same sky as used in MORX.  The spurious objects are gone, the POSS-I plate outlines are seen as before although broadened by USNO-B data, and overlaying these are the broad outlines of the overlapping POSS-II plates from USNO-B.  The dense spots at the NE corners of the POSS-I plates are a USNO-B artefact consisting of faint (and presumably spurious) objects of \textit{O}$\approx{21.1}$ \& \textit{E}$\approx{20.5}$.  Apart from that, the new optical background shows far better uniformity and greater depth (from the POSS-II data) compared with the previous.  This enables more optical matches to radio/X-ray sources whilst avoiding spurious matches, and also has remedied a problem with optical field solutions as is detailed in Section 4.

\begin{figure} 
\includegraphics[scale=0.45, angle=0]{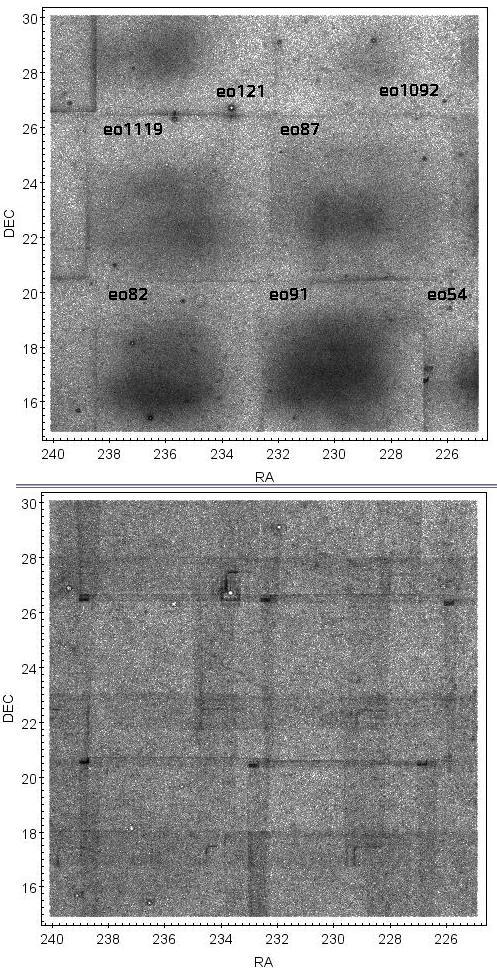} 
\caption{Comparative object densities of the optical backround, darker is denser, tile of sky is RA 15h-16h, Dec 15$^{\circ}$-30$^{\circ}$, old background (top) \& new (bottom).  SDSS data is excluded.  Top panel shows POSS-I plate names.} 
\end{figure}

110M additional optical sources distinct from the APM/USNO data were added from the SDSS, bringing the total optical data to 1\,043\,386\,316 objects.  Photometry flags identify the origin and other attributes of the optical data, see the ReadMe for the full index.

\section{Radio and X-ray input catalogues}

The input radio/X-ray catalogues used are all the large high-precision ones.  For radio sources there are the NRAO VLA Sky Survey catalogue (NVSS) \cite{NVSS} which covers the whole sky north of declination -40$^{\circ}$ in 1.4GHz to a completeness limit of 2.5mJy and positional uncertainty of $\approx$ 5 arcsec, the Faint Images of the Radio Sky at Twenty-cm survey catalogue (FIRST) \cite{FIRST} which is a 1.4GHz survey with a footprint subset of NVSS but with a completeness limit of 1mJy and positional uncertainty of just 1 arcsec, and the Sydney University Molonglo Sky Survey catalogue (SUMSS) \cite{SUMSS} which is analagous to NVSS but covers the sky south of declination -30$^{\circ}$ in 843MHz.  For X-ray sources there are 13 input catalogues from the \textit{Chandra}, \textit{XMM-Newton}, \textit{Swift}, and \textit{ROSAT} satellite surveys, these are: for \textit{Chandra} data, the Chandra ACIS source catalog (CXOG) \cite{CXOG}), the Chandra Source Catalog v1.1 (CXO) \cite{CXO}, and the XAssist Chandra source list (CXOX) \cite{XAssist}.  For \textit{XMM-Newton} data, catalogues used are the XMM-Newton DR6 (3XMM) \cite{XMMDR6}, the XAssist XMM-Newton source list (XMMX) \cite{XAssist}, the XMM-Newton DR3 (2XMM) \cite{XMMDR3}, and the XMM-Newton Slew survey release 1.6 (XMMSL) \cite{Slew}.  For \textit{ROSAT} data, catalogues used are the High Resolution Imager (HRI) \cite{PSPC}, Position Sensitive Proportional Counter (PSPC) \cite{PSPC}, the WGACAT (WGA) \cite{WGA} which covers the same data as PSPC but with different processing, and the revised all-sky survey (RASS2) \cite{RASS2} and some sources from the original all-sky survey (RASS1) \cite{RASS1} where they accord to other surveys' findings.  Lastly, \textit{Swift} data is taken from the Swift X-ray Point Source catalog (SXPS) \cite{SXPS}.  Counts of sources from these input catalogues are given in Table 2.

Some of these catalogues present data from the same operational survey and so overlap in large part.  The CXOG, CXOX and CXO catalogues share extensive \textit{Chandra} survey data, the 3XMM, XMMX and 2XMM share extensive \textit{XMM-Newton} survey data, and the PSPC and WGA catalogues share most \textit{ROSAT} PSPC survey data although processed very differently.  All these overlapping catalogues have been de-duplicated against their fellows so that any individual survey detection/source never appears twice in MORX under any guise.  In detail: CXO catalogues were de-duplicated against each other within an 8 arcsec radius of raw source positions, or 9 arcsec with similar flux.  XMM catalogues are the same but up to 12 arcsec with similar flux.  PSPC-WGA de-duplication was done within 22 arcsec radius, or 36 arcsec with similar flux.  These thresholds were identified via heavy tails over the background.  Supplementary de-duplication of associations across catalogues followed similar rules.  265K CXO duplicates, 334K XMM duplicates, and 87K PSPC-WGA duplicates were removed.        

In principle, the 2XMM catalogue should be entirely superseded by 3XMM, but in practice 3XMM lost some 18\,000 valid 2XMM sources; this is discussed in section 8.2 of Rosen et al. \shortcite{XMMDR6} which refers to them as ``missing 3XMM detections''.  Such a number of 3XMM X-ray sources would yield $\approx$6500 optical-X-ray associations.  Processing of the full 2XMM-DR3 data yields 15\,505 optical-X-ray associations which are unmatched to the 3XMM associations, with a confidence range of 40\%-100\%.  But the expectation of the ``missing 3XMM detections'' is $\approx$6500 associations, not 15\,505.  Therefore I've replicated that by requiring that the extra 2XMM associations should be of confidence$\ge$80\%, which yields 4979 robust 2XMM associations unmatched to the 3XMM associations in MORX.  In this way those ``missing 3XMM detections'' have been retrieved.

The recent RASS2 \cite{RASS2} release involved extensive re-processing of the RASS1 \cite{RASS1} data, and the outcome is a significantly different catalogue with many sources shifting over an arcminute away.  I've taken RASS2 as superseding RASS1, but processing revealed 547 matched RASS1-RASS2 sources where the RASS1 association accords with other X-ray associations to the same optical object, while the RASS2 partner does not (and usually associates to nothing).  In those 547 cases I have retained the original RASS1 source and dropped its RASS2 partner, and in all other cases I use the RASS2 data only.

\section{Optical Field Solutions}

The task in MORX is to gauge the likelihood that a particular optical object is the true source of a radio or X-ray detection, and a full discussion of this is given in the QORG paper, but of course astrometric co-positionality is a powerful indicator.  To best evaluate this, one must correct for systematic errors on each side, that is, align the radio/X-ray observation field/sequence with the optical data -- this is called the ``optical field solution'' (OFS).  MORX uses only detections bearing raw astrometry and field identifications which are provided by all the input catalogues listed above.  Some input catalogues also offer ``corrected'' or ``master'' source positions for which OFSs or detection stacking have been performed, or optical sources selected, but those are not utilized here as MORX provides independent automated optical selections (but Table 3 shows a comparison which is also discussed below).  One small but prominent catalogue, the Chandra Multiwavelength Project X-ray Point Source Catalog \cite{CXOMP}, is not included in MORX because they did not present their raw astrometry or field identifications.  The loss is small because almost all their sources are also covered by the CXOG, CXOX or CXO catalogues. 
  
MORX performs OFSs via a standard algorithm operating on the raw detection positions of each radio/X-ray field, sliding them over the near optical background to find best fits for X-rays and radio cores.  All fields are processed thusly, and detections from those fields with the highest confidence OFSs are kept in preference to others at the time of de-duplication.  Note that some radio/X-ray fields have too few sources, or too few optical matches, to find an OFS; in such cases, the original astrometry is used, but at the subsequent de-duplication it is the detections from fields with successful OFSs which are the preferred ones to retain.   
       
Table 2 shows the typical and maximum astrometric shifts of the OFSs found for each input catalogue.  \textit{ROSAT} fields from the 1990's require the largest shifts, and furthermore their all-sky catalog (RASS) has fields of calculational convenience, not observational fields (RASS observational sequences were long arcs across the sky), so the alignment method has reduced utility.  \textit{Chandra} fields have excellent astrometric accuracy (so good that their authors don't bother with OFSs, but MORX does calculate them), followed closely by \textit{XMM-Newton} fields.  About 30\% of the SXPS detections in MORX are stacked (the ones having 4-digit field IDs), thus are multi-field accumulations, so the alignment method is less effective as with RASS.  \textit{ROSAT} field OFSs sometimes differ from that displayed in the QORG \shortcite{QORG} documentation, because those legacy OFS calculations were hampered by an optical background with many spurious one-colour objects, now removed as discussed in Section 2.  The input radio data has excellent astrometric accuracy, usually better than the MORX optical data has, but that optical data serves as the reference point for the calculations so the radio data must nontheless be aligned to it, even if equal errors (of the order of 1 arcsecond) are introduced to align it.  

The astrometric shifts of the OFS for each individual radio/X-ray field are given in the right-hand columns of MORX; for objects with more than one radio/X-ray association, the OFS of the highest-likelihood association is presented, except that an X-ray field with a successful OFS is always preferred.  Throughout the full million lines of data, all radio/X-ray fields with successful OFS are displayed at least once; for those displayed many times, the information is identical each time.  As a comparison, Table 3 shows the counts of MORX OFS astrometric shifts (in arcseconds) for the 3XMM X-ray catalogue, along with the same as presented by 3XMM which is the only input catalogue to present OFSs in a similar vein as MORX.  The last column counts the offsets between each side's OFSs, that is, how much they disagree; this shows 97.8\% agreement within 2 arcsec.

When the radio/X-ray fields have been aligned thusly to the optical background, then the likelihood algorithms can be applied.

\begin{table*} 
\scriptsize	 
\caption{Summary of Optical Field Solutions (OFS) of the Source Catalogues (shifts are in arcseconds)}
\begin{tabular}{@{\hspace{0pt}}l@{\hspace{0pt}}r@{\hspace{6pt}}r@{\hspace{6pt}}rrrrrccccr}
\hline 
           & \#      & \#         & \#       &       &        &       &       & EW    & NS    & EW      & NS      & \#  \\
Source     & unique  & matched    & matched  & mean  & median & max$^{*}$ & min & mean & mean & shift$^{\dagger}$ & shift$^{\ddagger}$ & matched \\
Catalog(s) & sources & to optical & with OFS & shift & shift  & shift & shift & shift & shift & min/max & min/max & w/o OFS \\
\hline 
CXO v1.1   &  106347 &  41012 &  37832 & 0.52 & 0.00 &  3.36 & 0.00 & -0.02 & +0.09 &  -2.7/+2.3  &  -2/+3  & 3180 \\
CXOG       &  217607 &  69833 &  62120 & 0.42 & 0.00 &  2.80 & 0.00 & -0.01 & +0.09 &  -2.8/+2.0  &  -2/+2  & 7713 \\
CXOX       &  297800 &  71671 &  64378 & 0.49 & 0.00 &  2.97 & 0.00 & -0.01 & +0.10 &  -2.8/+2.5  &  -2/+2  & 7293 \\
{\it total Chandra} & & 91666 &  81007 & 0.46 & 0.00 &  3.36 & 0.00 & -0.01 & +0.09 &  -2.8/+2.5  &  -2/+3  & 10659 \\
3XMM-DR6   &  463218 & 174091 & 141216 & 1.18 & 1.00 &  4.47 & 0.00 & -0.03 & +0.16 &  -4.0/+4.2  &  -4/+4  & 32875 \\
2XMM-DR3   &  250026 &  98409 &  82417 & 1.49 & 1.35 &  5.59 & 0.00 & +0.22 & +0.17 &  -4.7/+4.9  &  -3/+4  & 15992 \\
XMMX       &  239240 &  50441 &  40845 & 1.56 & 1.41 &  5.39 & 0.00 &  0.00 & +0.35 &  -5.0/+5.0  &  -4/+5  & 9596 \\
{\it total XMM-Newton} & & 194683 & 147849 & 1.25 & 1.10 & 5.59 & 0.00 & -0.01 & +0.18 & -5.0/+5.0 & -4/+5  & 46834 \\
XMM Slew v1.6       &  18280 & 10108 &  7681 & 2.23 & 2.00 & 12.12 & 0.00 & +0.14 & +0.27 &  -9.0/+9.9  &  -6/+7  & 2427 \\
\textit{ROSAT} HRI  &  54689 & 19539 & 14012 & 2.68 & 2.29 & 16.03 & 0.00 & -0.22 & -0.55 & -15.0/+10.7 & -16/+9  & 5527 \\
\textit{ROSAT} PSPC & 101402 & 43017 & 29979 & 2.87 & 2.62 & 14.14 & 0.00 & -0.34 & -0.01 & -12.7/+8.8  & -10/+11 & 13038 \\
\textit{ROSAT} WGA  &  76751 & 28172 & 16394 & 6.19 & 5.97 & 22.38 & 0.00 & +1.92 & -1.10 & -16.3/+20.9 & -21/+19 & 11778 \\
\textit{ROSAT} RASS & 129194 & 51958 & 30931 & 2.98 & 2.91 & 11.24 & 0.00 & -0.24 & -0.73 &  -8.1/+7.1  & -11/+9  & 21027 \\
\textit{Swift} SXPS &  88294 & 48190 & 40114 & 1.91 & 1.70 & 10.82 & 0.00 & +0.10 & +0.29 &  -9.0/+9.0  &  -8/+9  & 8076 \\
FIRST core &  939885 & 247133 & 206497 & 0.63 & 0.90 &  2.97 & 0.00 & -0.02 & +0.12 &  -2.8/+2.3  &  -2/+2  & 40636 \\
NVSS core  & 1810664 & 385057 & 342198 & 0.44 & 0.00 &  2.27 & 0.00 & -0.03 & +0.15 &  -1.9/+1.1  &  -1/+2  & 42859 \\
SUMSS core &  259896 &  68178 &  49597 & 0.94 & 1.00 &  3.35 & 0.00 & +0.41 & +0.27 &  -2.2/+3.2  &  -2/+2  & 18581 \\
\hline
\multicolumn{13}{@{\hspace{0pt}}l}{$^{*}$ Published raw positional uncertainties for these catalogues are: \textit{Chandra}: 80\% within 2 arcsec, 97\% within 4 arcsec;} \\
\multicolumn{13}{@{\hspace{0pt}}l}{\quad \textit{XMM-Newton}: aspect mean 2 arcsec, max 6 arcsec;  XMM Slew: 68\% within 8 arcsec, 90\% within 17 arcsec;} \\ 
\multicolumn{13}{@{\hspace{0pt}}l}{\quad \textit{Swift}: aspect median 3.5 arcsec, max 15 arcsec;} \\
\multicolumn{13}{@{\hspace{0pt}}l}{\quad \textit{ROSAT} PSPC: nominal 15 arcsec, max 25 arcsec; HRI: in practice, same as PSPC (used same equipment), nominal was 5 arcsec;} \\
\multicolumn{13}{@{\hspace{0pt}}l}{\quad WGA: add 10 to PSPC values (early solution missed subsequent aspect corrections); RASS: nominal 30 arcsec, can be much higher;} \\
\multicolumn{13}{@{\hspace{0pt}}l}{\quad FIRST / NVSS / SUMSS: ground-based astrometry taken as true, adjustments onto my optical background with max 2 arcsec error.} \\
\multicolumn{13}{@{\hspace{0pt}}l}{$^{\dagger}$ EW shifts are evaluated in integer RA arcsec, i.e., 1/3600$^{th}$ of a RA degree, converted here to true arcseconds.} \\
\multicolumn{13}{@{\hspace{0pt}}l}{$^{\ddagger}$ NS shifts are evaluated in integer arcseconds.} \\
\end{tabular}
\end{table*}

\begin{table} 
\scriptsize	 
\caption{OFS-driven shifts of 3XMM sources (in arcseconds)}
\begin{center} 
\begin{tabular}{crrr}
\hline 
 arcsec   & MORX & 3XMM & MORX-3XMM  \\
(rounded) &  OFS &  OFS & OFS offset \\
\hline 
0 & 182562 &  77538 &  99928  \\
1 & 193757 & 275617 & 310900  \\
2 &  70418 &  89948 &  42414  \\
3 &  14683 &  16981 &   7264  \\
4 &   1645 &   2701 &   2399  \\
5 &    153 &    432 &    240  \\
6 &      0 &      0 &     73  \\
7 &      0 &      1 &      0  \\
\hline
\end{tabular}
\end{center}
\end{table}

\section{Association Likelihoods of Radio/X-ray Sources}

Likelihood of X-ray or core radio association to optical objects is calculated by comparing optical areal densities on the sky as follows.  A detailed explanation is given in the QORG paper appendix A section 2, but salient points and data changes are presented here.

Firstly, the optical sky (excluding SDSS data, see Section 2) is partitioned into 12 density classes from low-density to high-density.  Table 4 shows the sky area and counts of each partition, plus how many MORX objects hail from there; Figure 3 maps them onto the sky by colour.  These density partitions are essential because they serve as the baseline against which to calculate overabundance, e.g., a 2 arcsecond offset is more significant in sparse sky than in dense sky.  The partitions were chosen for large counts and density increase steps of 20\%-25\% so that the baseline density should be correct within $\approx$10\% for each object, and the counts and areas show good uniformity.  Also, sky object mix changes as density increases, because Galactic stars increase while extragalactic objects stay constant or decrease due to extinction.  These 12 sky partitions are hereafter used in isolation from eachother; each serves as the full sky baseline \textit{a.k.a.} optical background for its own objects.     

\begin{figure*} 
\includegraphics[scale=0.5, angle=0]{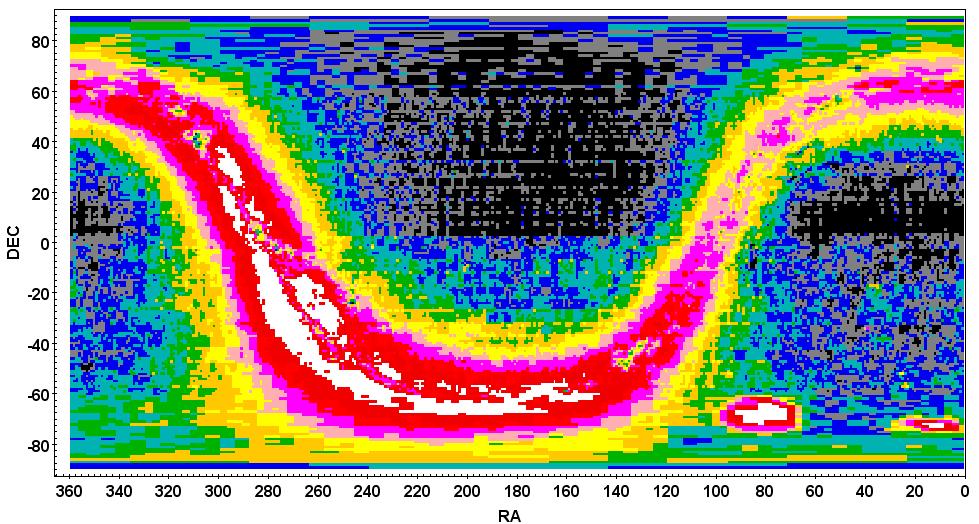} \\
\caption{Optical sky density classes mapped on the sky -- see Table 4 for index of colours.  Granularity is 1 RA degree x 1 Dec degree, except near the poles where RA granularity stretches.  Density boundary near the equator is due to deep APM scans of South-sky UKST plates (refer Section 2).  SDSS data is not used.} 
\end{figure*}

\begin{table} 
\scriptsize	 
\caption{Optical Sky Density Classes}
\begin{center} 
\begin{tabular}{@{\hspace{0pt}}c@{\hspace{4pt}}rr@{\hspace{4pt}}rl@{\hspace{2pt}}r}
\hline 
 density &  sky   &  count of     &  mean      & colour   & count of  \\
class ID &  area  &  optical      &  density   & on       & MORX    \\
 (=min)  & sq deg &  sources      &  (ct/area) & Figure 3 & objects \\
\hline 
 LOW  &  5845.7  &  39\,600\,709  &   6\,774  & black       &  216\,230  \\
  8K  &  5766.9  &  51\,832\,781  &   8\,988  & grey        &  201\,069  \\
 10K  &  5448.1  &  59\,760\,117  &  10\,969  & blue        &  160\,875  \\
 12K  &  5554.7  &  74\,351\,027  &  13\,385  & light blue  &  143\,658  \\
 15K  &  3366.3  &  55\,117\,210  &  16\,373  & green       &   65\,209  \\
 18K  &  2990.7  &  59\,185\,742  &  19\,790  & dark yellow &   54\,585  \\
 22K  &  2695.6  &  66\,938\,137  &  24\,833  & yellow      &   35\,110  \\
 28K  &  2420.3  &  75\,796\,930  &  31\,317  & light pink  &   26\,916  \\
 35K  &  2379.8  &  94\,738\,890  &  39\,810  & dark pink   &   25\,692  \\
 45K  &  1932.1  &  96\,286\,566  &  49\,836  & bright red  &   23\,486  \\
 55K  &  1533.3  &  95\,255\,664  &  62\,126  & dark red    &   19\,571  \\
 80K  &  1319.6  & 164\,361\,070  & 124\,552  & white       &   30\,454  \\
\hline 
 ALL  & 41253.0  & 933\,224\,843  &  22\,622  &             & 1\,002\,855 \\
\hline
\end{tabular}
\end{center}
\end{table}

Next all the optical objects are classified by red-blue colour and PSF.  PSF has but 3 values being stellar, fuzzy, or unknown, so across red \& blue there are 9 combinations.  Red-blue colour is binned by \textit{B-R} steps of 0.3 into 17 discrete values of -0.6 to 4.2, plus 2 catch-all endpoints, plus 2 placeholders for red-only and blue-only objects.  Thus the optical objects in each sky density partition are further subdivided into these 189 (=9x21) subsets, although the ``unknown'' PSF bin has larger numbers because in practice it totals all PSFs (because ``unknown PSF'' is not a physical state).  These subsets supply the denominators for density calculations.      

Now the following procedure is done independently for each input radio/X-ray catalogue:  Optical neighbours are found for all radio/X-ray detections and are binned by astrometric offset (in integer arcseconds) to those detections -- each offset bin has a sky area of an annular ring of arcsecond ring width -- these are totalled over all detections for each optical class and sky density partition (as defined above in this section).  The counts and total sky areas are then compared to the denominators of their optical background; the outcome is an areal density ratio.  Twice the background optical density means that we expect that half of that bin's optical objects are causally associated to their radio/X-ray detections, and so on.  The QORG paper elaborates on this extensively.   

Testing against SDSS optical objects is done separately against the same non-SDSS optical background, but the fainter SDSS objects actually comprise a denser sky than that background, so to compensate for that, the output areal densities are halved as an approximation.  

The areal density ratio (or just ``density'') is a number like e.g., 10 which represents an overdensity of 9 plus the background of 1.  This is written in MORX as the confidence percentage \textit{(density-1)/density}, so a density of 10 is displayed in MORX as confidence = 90.0\%.

\section{Double Radio Lobes}

Identification of double radio lobes is a task orthogonal to the others here.  The MORX algorithm uses data characteristics of the input FIRST, NVSS and SUMSS radio catalogues to heuristically data-mine out the true lobes and which optical centroid is the likely true emitter.  Refer to the QORG appendix A section 4 for the (excruciating) details, but one improvement to the QORG algorithm will be described.

In general, the task is to find eligible radio detections, i.e., orphan detections without optical sources, and see if a compatible pair of such are arrayed symmetrically around a suitable optical object.  Detection ellipse geometry, triple configuration, and optical photometry are all considered in rating the lobe-source candidate.  When the rating exceeds 40\% confidence -- a measure heuristically calibrated against hundreds of radio cut-out images -- then the optical candidate is entered into MORX as the source of the adjudged lobes.  

If no core radio association is present, then the lobe association confidence is displayed in the radio association confidence column for that object.  However, if there is a core radio association, then the working assumption is that the radio core is the true source of the lobes, and thus the confidence that the core-lobe combination are associated to that optical object is simply the confidence of the core radio association.  This is because if the core radio association were false, for example if it were actually associated to an unseen background object, then also the lobe association would be.  Thus the confidence of association of double lobes is never greater than the confidence of association of an accompanying radio core.

In total, 27\,378 double lobe associations are presented in MORX, 5661 with radio cores and 21\,717 without.  Of these, 3849 are for classified quasars (half of which have radio cores), 7426 galaxies, and 16\,036 are anonymous.  Also, 67 classified stars are so presented which is of course very unlikely to be true.  Whether the cause is an unseen background source (always a hazard), misclassification as a star, or false lobes, is a matter for individual investigation; they are left in MORX because of its automated nature.          

A recent change to the double lobe algorithm is the projected placement of the core along the lobe midline, previously taken to be the midpoint of that midline.  Not taken into account by this model were long-standing findings such as by \cite{McCarthy} that where lobes are asymmetrical, the shorter lobe is usually the brighter of the two.  This has now been worked into the model as an adjustment on the distpct ($\delta$, see QORG appendix A.4) dependent on the SNRpct.  The outcome is better reliability of the optical centroid selection where there is no core radio detection.  As to the cause of the short lobe being brighter, the consensus is that it's an IGM environmental effect, although from viewing many of these my own assessment is that lobe projection should play some role; that is, the shorter lobe may be pointing more into our line of sight so that we see it as shorter and brighter as a projection effect.

\section{Probabilistic Classification as QSO, Galaxy or Star}

All MORX objects show calculated percentage likelihoods that the object is a QSO, galaxy, star, or that the presented radio/X-ray association is erroneous.  These likelihoods are obtained by binning all associations by optical PSF, red-blue colour, radio/X-ray flux to optical magnitude comparison, and astrometric offset of the detection to the optical.  Of the million objects in this catalogue, 321\,024 are already classified (including photometric) in the literature as QSOs, galaxies or stars, so for each combination of the 4 bins, its classified members summate to QSO, galaxy and star percentages which are taken as being the likelihoods of those classifications, and those likelihoods are then allocated to the anonymous members of that combined bin; the classified members also display those likelihoods, for user inspection.

However, each object's radio/X-ray association is marked by a confidence percentage.  The corresponding error likelihood is simply the complement of that confidence percentage.  If the presented association were false, then the above classification likelihoods would not apply, thus the QSO, galaxy and star likelihoods must be correspondingly reduced to accomodate the error likelihood such that all 4 likelihoods sum to 100\%.  
   
The error likelihood is the only place in MORX where the radio association confidence and X-ray association confidence are combined.  For objects with both radio and X-ray associations, the respective optical densities are simply overlaid, that is, added (with the extra background of 1 subtracted), and the confidence of association then calculated from that joint density figure.  That joint confidence of association is subtracted from 100\% to yield the error likelihood for that object.

\section{Legacy data imported from ARXA}

The Atlas of Radio/X-ray Associations (ARXA: Flesch \shortcite{ARXA}) was the predecessor to MORX; it presented 602\,570 radio/X-ray associations.  The present MORX processing recaptured 466,008 of those, meaning 136\,562 ARXA objects were dropped, an unexpectedly high number.  Investigation of the dropped data found that the inclusion of USNO-B optical data was the prime mover in this, often changing optical photometry to something less quasar-like or galaxy-like, thus shrinking the associative radius; or changing the optical position to something farther from the radio/X-ray source position; and sometimes new USNO-B optical objects ``stole'' the association from the previous owner.  Some optical field shifts changed.  Many one-colour optical data were dropped as described in Section 2, causing any associations to those optical data to be dropped also.  In some cases the relevant radio or X-ray detection was dropped from an updated input catalogue.  Also, Section 3 described the removal of RASS1 as an input catalogue, but 39\,188 ARXA associations were to RASS1 sources only.  Of those, 17\,210 matched to new RASS2 sources in MORX, and 21\,978 were unmatched and so dropped.  And double lobe calculations changed when radio sources, previously unassociated and so eligible as radio lobes, became associated to new USNO-B optical data, so making them unavailable to the lobe algorithm; 7543 ARXA double lobes were thus dropped.    

However, spot checks also showed many attractive candidate associations amongst the orphaned ARXA data, so I thought it best to identify a desirable subset to salvage.  This subset consists of those still-valid radio/X-ray detections which are associated to good-photometry (both red \& blue present) optical objects, or with confidence of $>$60\%, and which are located $>$10 arcsec away from all MORX sources and associations.  Added to this subset were 474 FIRST radio detections which were dropped from the latest FIRST catalogue because their fluxes were recalculated as less than the FIRST 1mJy threshold; these sources generally look true on the FIRST cutout server and, where so, comprise authentic associations.     

This set of salvaged ARXA data amounts to 39\,157 associations which have been added into MORX, thus raising the total number of ARXA objects retained in MORX to 505\,165.  The salvaged data are mostly NVSS associations at offsets of 1-10 arcsec from the optical, but also 3112 attractive FIRST associations and 7605 X-ray associations of mixed quality.  It's difficult to quality check the NVSS associations because of the comparatively low resolution of NVSS detections, so they are presented in the hope that their inclusion will be useful.  This data is flagged with an ``x'' in the photometry comment column, and with the text ``from ARXA'' in the field/sequence column.

\section{Catalogue Layout}

The catalogue is written as one line per optical object, and sample lines from the catalogue are displayed in Table 1; the block of lines are wrapped, with the left half displayed on top of the right half.  The optical object is named where it is classified; else, if anonymous, the sexagesimal J2000 location is displayed as a convenience to the user.  Other explanations are given throughout this paper, and the ReadMe gives full details and indexes to the columns.  

There is one X-ray detections column for each of the X-ray satellites, being \textit{XMM-Newton}, \textit{ROSAT}, \textit{Chandra} and \textit{Swift}; the column orders are simply by how many sources each contributes to MORX.  There are two radio detection columns, one for NVSS and the other displaying both FIRST and SUMSS sources because their footprints don't overlap; it's expected that users will be aware of the difference between the high-resolution FIRST survey and SUMSS which is analagous to NVSS.  Just one set of double lobe associations are presented, for brevity.  There are 873 lines for which both FIRST and NVSS double lobes were available; the lobes with highest confidence were selected in those cases.
   
Optical objects and radio/X-ray sources each appear once only in MORX.  Radio/X-ray sources are identified by name, and the user can use that name to consult the original radio/X-ray catalogue for any additional information on that source.  The Chandra ACIS source catalog \cite{CXOG} presents detection names prefaced by ``CXOGSG''; this is cropped here to ``CXOG'' for brevity.  Similarly, the XMM Slew detection prefix ``XMMSL1'' is here cropped to ``XMMSL''.  NVSS radio detection names are presented in the same \textsl{Jhhmmss.s+ddmmss} format as the other surveys, to display the astrometry.  The NVSS-recommended form \textsl{NVSS Jhhmmss+ddmmss} can be recovered by simply removing the time decimal.
   
The two right-hand columns give references for classified object names and redshifts.  The classification (e.g., that the object is a quasar) can come from either.  The references are indexed with full citations in the accompanying file ``MORX-references.txt'' which is included in the MORX zip file.

\section{Miscellaneous Notes}
 
There is some minor de-duplication of low-confidence associations done across input catalogues at the end of processing, up to 8 arcseconds offset if very low confidence.  Radio/X-ray sources associated to large galaxies or bright stars or high-proper-motion stars are de-duplicated by optical object name, so that the detection names can show quite different astrometry, but they should be correctly allocated to the large (or fast-moving) object on the sky.  

Radio and X-ray associations are processed separately without any reference to each other at any time.  In principle, MORX could have been published as two catalogues, one for radio associations and one for X-ray associations.  Where any optical object has both radio and X-ray associations, it earned each one independently.

\section{Conclusion}

This catalogue compiles and optically aligns {\it Chandra}, {\it XMM-Newton}, {\it ROSAT} and {\it Swift} X-ray sources, and NVSS, FIRST and SUMSS radio sources, and presents those calculated to be associated to optical objects.  Identifications are included to present an informative map to support pointed investigations.  Its total count is 1\,002\,855 optical objects with 40\%-100\% likelihood of radio/X-ray associations including double radio lobes.

\begin{acknowledgements}
Thanks to Phil Helbig for spurring me on to make this final catalogue.  Thanks to the anonymous referee for numerous suggestions which have helped to clarify this paper.  Thanks for users' kind comments on past work which made it all feel worthwhile.  This work was not funded.
\end{acknowledgements}

\end{document}